\documentclass[fleqn,twoside]{article}
\usepackage{espcrc2}

\usepackage{graphicx}
\usepackage[figuresright]{rotating}
\usepackage{amssymb}

\newcommand{\lra}{\longrightarrow}
\newcommand{\beq}{\begin{equation}}
\newcommand{\eeq}{\end{equation}}
\newcommand{\etal}{{\em et al.}}
\newcommand{\cO}{\cal{O}}

\newcommand{\AmS}{{\protect\the\textfont2
  A\kern-.1667em\lower.5ex\hbox{M}\kern-.125emS}}

\begin{document}

\thispagestyle{empty}

\title{
\vspace{-1.8cm}
\hfill \rm \null \hfill
 \hbox{\normalsize ADP-01-58/T490} \\
\vspace{+1.3cm}
Cooling for instantons and the Wrath of Nahm}

\author{S. Bilson-Thompson, F. D. R. Bonnet, D. B. Leinweber, and A. G. Williams 
        \address{Centre for the Subatomic Structure of Matter and Department of 
         Physics and Mathematical Physics, \\  
        University of Adelaide, Adelaide 5005, Australia}}

\begin{abstract}
The dynamics of instantons and anti-instantons in lattice QCD can be studied by analysing 
the action and topological charge of configurations as they approach a self-dual or 
anti-self-dual state, i.e. a state in which $S/S_0=|Q|$. We use cooling to reveal the 
semi-classical structure of the configurations we study. Improved actions which eliminate 
discretization errors up to and including ${\cO}(a^4)$ are used to stabilise instantons as 
we cool for several thousand sweeps. An analogously 
improved lattice version of the continuum field-strength tensor is used to construct a 
topological charge free from ${\cO}(a^4)$ discretization errors. Values of the 
action and topological charge obtained with these improved operators approach 
mutually-consistent integer values to within a few parts in $10^4$ after several hundred 
cooling sweeps. Analysis of configurations with $|Q| \approx 1$ and $|Q| \approx 2$ supports 
the hypothesis that a self-dual $|Q|=1$ configuration cannot exist on the 4-torus.

\vspace{1pc}
\end{abstract}

\maketitle

\section{INSTANTONS}

In 1975 Belavin {\etal} \cite{BelPolSchTyu} determined that the vacuum of QCD contained 
non-trivial local minima of the Euclidean action,
\beq 
S_E = \frac{1}{2g^2} \int d^4x {\mathrm{Tr}}F_{\mu\nu}F_{\mu\nu}.
\eeq
Since this action is required to be finite in the case of semi-classical solutions 
(i.e. solutions obtained by minimizing the action) the definition of the field-strength 
tensor
\beq
F_{\mu\nu}=\partial_{\mu}A_{\nu} - \partial_{\nu}A_{\mu} + [A_{\mu},A_{\nu}]
\eeq
implies that 
\beq
A_{\mu} \longrightarrow U^{-1}\partial_{\mu}U
\eeq
as $|x|$ goes to infinity. The gauge transformation $U$ is interpreted as a mapping
from $S^3 \longrightarrow SU(2)$, that is, a mapping from the sphere at spacial infinity
to the gauge group $SU(2)$ of the transformation $U$ ($SU(2)$ results can be readily extended 
to $SU(3)$). The mapping $U$ can then be shown to have a winding number given by the equation
\beq
n = \frac{1}{16 \pi^2}\int d^4x {\mathrm{Tr}}F_{\mu \nu} \widetilde{F}_{\mu \nu}
\eeq
and hence we can associate a winding number with each vacuum state. Instantons are
interpreted as tunnelling amplitudes between vacuum states with different winding numbers,
and their topological charge is given by the difference between the winding numbers of the 
states that they connect. Further important properties of (anti-)instantons are 
spherical symmetry in the four dimensions of Euclidean spacetime, topological charge of $\pm 1$,
and an associated action of $S_0 = 8\pi^2/g^2$ which is independent of the radius of the instanton.

\section{CONNECTING THE LATTICE TO THE CONTINUUM}

The lattice is a 4-dimensional approximation to Euclidean spacetime. This means that physics
on the lattice should ideally approach physics in the continuum if we can take the limit of 
infinite lattice volume and infinitessimal lattice spacing ($V \lra \infty$ and $a \lra 0$).
Unfortunately reducing the lattice spacing indefinitely is not practical, nor is increasing 
the total lattice volume, since these approaches to the continuum limit rapidly increase the 
time taken to perform computations.\\
We have chosen to remove the non-physical `edge' of the lattice (simulating an infinite volume)
by defining periodic untwisted boundary conditions on our lattices. We are therefore performing
numerical QCD on a 4-toroidal mesh of points. The discretization errors introduced by the finite 
lattice spacing are reduced algebraically by Symanzic improvement \cite{Symanzic} and non-classical 
errors arising from the self-couplings of the gluon fields are dealt with by the use of 
mean-field improvement \cite{LePage}.\\
Since the lattice we use is topologically equivalent to a 4-torus, we may consider the 
consequences of the Nahm transform \cite{GonzArrPena} for the stability of the configurations 
we study. The Nahm transform is a duality mapping which interchanges between an $SU(N)$ configuration with
topological charge $Q$ on the torus and an $SU(Q)$ configuration with topological charge $N$ on
the dual torus. Since there are no instanton solutions in $U(1)$, the Nahm transform implies
\cite{Taubes}\cite{Schenk} that there can be no self-dual $|Q|=1$ configurations on the torus.

\section{THE LATTICE ACTION AND TOPOLOGICAL CHARGE}

\subsection{The Wilson Action}
The Yang-Mills action on the lattice \cite{Wilson} is
\beq
S_\mathrm{Wil} = \beta \sum_{x} \sum_{\mu<\nu}
	\left[1 - \frac{1}{N}\left(Re\mathrm{Tr}\,W_{\mu \nu}^{(1 \times 1)}(x)\right)\right]
\label{eq:WilsonAction}
\eeq
where $W_{\mu \nu}^{(m \times n)}$ is the $m \times n$ Wilson loop in the \mbox{$\mu-\nu$} 
plane. When we expand the plaquette $W_{\mu \nu}^{(1 \times 1)}$ around the point $x$ and set 
$\beta = 6/g^2$ it can be easily shown that (\ref{eq:WilsonAction}) approaches the continuum 
Yang-Mills action, plus ${\cO}(a^2)$ corrections
\beq
S_\mathrm{Wil} \lra \frac{1}{2} \int d^4x {\mathrm{Tr}}F_{\mu \nu}F_{\mu \nu} + {\cO}(a^2) 
\eeq
in the continuum limit $a \lra 0$.

\subsection{Improving the Action}
We can eliminate ${\cO}(a^2)$ errors by constructing the Wilson action from a linear 
combination of the plaquette and the average of the $1 \times 2$ and $2 \times 1$ 
rectangular Wilson loops. We may of course choose to employ other loops to eliminate 
higher-order error terms. DeForcrand {\etal} \cite{DeForc} have previously constructed an
action free from ${\cO}(a^2)$ and ${\cO}(a^4)$ errors by using the five planar Wilson loops 
\begin{itemize}
	\item $L^{(1,1)} = W_{\mu \nu}^{(1 \times 1)}$
	\item $L^{(2,2)} = W_{\mu \nu}^{(2 \times 2)}$
	\item $L^{(1,2)} = \frac{1}{2}\left(W_{\mu \nu}^{(1 \times 2)} 
                                          + W_{\mu \nu}^{(2 \times 1)}\right)$
	\item $L^{(1,3)} = \frac{1}{2}\left(W_{\mu \nu}^{(1 \times 3)} 
                                          + W_{\mu \nu}^{(3 \times 1)}\right)$.
	\item $L^{(3,3)} = W_{\mu \nu}^{(3 \times 3)}$
\end{itemize}
A general improved action can then be written in the form
\begin{eqnarray}
S_\mathrm{Imp} & = & c_1 S(L^{(1,1)}) + c_2 S(L^{(2,2)}) + c_3 S(L^{(1,2)}) \nonumber \\
               &   & c_4 S(L^{(1,3)}) + c_5 S(L^{(3,3)}).
\end{eqnarray}
where the $c_1,...,c_5$ are improvement factors which take the values 
\begin{eqnarray}
c_1 & = & (19 - 55 c_5 ) / 9 \nonumber \\
c_2 & = & (1 - 64 c_5) / 9 \nonumber \\
c_3 & = & (640 c_5 - 64 ) / 45 \nonumber \\
c_4 & = & 1/5 - 2 c_5 \nonumber 
\end{eqnarray}
and $c_5$ is a free parameter which we can use to ``tune'' the action. By setting $c_5=0$
we create a 4-loop improved action, denoted 4LIS for short, and by setting $c_5=1/10$ 
we eliminate the contribution of $S(L^{(1,2)})$ and $S(L^{(1,3)})$, thereby creating a
3-loop improved action, denoted 3LIS for short. For the purposes of investigating a 5LIS
we choose to follow DeForcrand {\etal} and set $c_5=1/20$, midway between the 3LIS 
and 4LIS values. It is important to note that the 3LIS and 4LIS are just special cases 
of the general 5LIS, and all choices of $c_5$ are free from discretization errors up to 
and including ${\cO}(a^4)$, however their ${\cO}(a^6)$ errors will be different, 
leading to slightly different results.\\ 

\subsection{Topological Charge}
On the lattice the topological charge is calculated as the sum of the local topological 
charge density over all lattice sites,
\beq
Q =\sum_{x} q(x) 
  =\frac{1}{32\pi^2}\sum_{x}\epsilon_{\mu\nu\rho\sigma}{\mathrm{Tr}}F_{\mu\nu}F_{\rho\sigma}.
\label{TopQDefn}
\eeq
If we have a dilute instanton gas with $n_I$ instantons and $n_A$ anti-instantons the total
topological charge of the configuration will be 
\beq
Q = n_I - n_A.
\eeq
We hence desire the topological charge calculated on the lattice to approach integer
values as the configuration being studied is cooled towards a dilute instanton gas 
condition. Unfortunately discretization errors tend to lead to non-integer values of 
$Q$. Fortunately, however, we can improve the topological charge.

\subsection{The Improved Field-Strength Tensor}
To improve the topological charge we have chosen to improve the field-strength tensor 
$F_{\mu\nu}$ directly, by analogy with the improvement of the action, and substitute 
this into Eq.~(\ref{TopQDefn}), the definition of the topological charge.\\
Consider the full expansion of the plaquette
\begin{eqnarray}
W^{(1 \times 1)}_{\mu \nu} & = & e^{ig\oint A dx}  \nonumber \\
       & = & 1+ig\oint A dx-\frac{g^2}{2}(\oint A dx)^2 \nonumber \\ 
       & & +{\cO}(g^3) \nonumber \\
       & = & 1+ig\left[a^2F_{\mu\nu}+{\cO}(a^4)\right]-\frac{g^2a^4}{2}F_{\mu\nu}^2 \nonumber \\ 
       & &      +{\cO}(a^6,g^3).
\end{eqnarray}
We wish to extract the second term (within the square parentheses) on the last line. 
This may be achieved by making the following construction;
\begin{eqnarray}
W^{(1 \times 1)}_{\mu \nu}     & = & 
 	1+ig\oint A dx -\frac{g^2}{2}(\oint A dx)^2 \nonumber \\ 
	& & + {\cal O}(g^3), \nonumber \\ 
W^{(1 \times 1){\dagger}}_{\mu \nu} & = & 
        1-ig\oint A dx -\frac{g^2}{2}(\oint A dx)^2 \nonumber \\ 
        & &  + {\cal O}(g^3).                                                         
\end{eqnarray}
Writing 
\beq
{\cal A}=W^{(1 \times 1)}_{\mu \nu} - W^{(1 \times 1){\dagger}}_{\mu \nu}
\eeq
we may obtain a term proportional to the field strength
\begin{eqnarray}
\Rightarrow \frac{-i}{2}\left({\cal A}-\frac{1}{3}\mathrm{Tr}{\cal A}\right)
	 & = & g\oint A dx + {\cO}(g^3) \nonumber \\ 
         & = & ga^2F_{\mu\nu} + {\cO}(ga^4). \nonumber \\ 
         &   & 
\end{eqnarray}
Notice that in order to enforce the tracelessness of the Gell-Mann matrices we
have subtracted one-third of the trace of ${\cal A}$.\\
To construct an improved field strength tensor we utilise the 
same five planar Wilson loops that are used in the construction of the 
improved action, but in this case we calculate $W^{(m\times n)}_{\mu\nu}$ from the 
clover average of four $m \times n$ Wilson loops in the $\mu-\nu$ plane. The improved 
field strength tensor can then be written
\begin{eqnarray}
F_{\mu\nu(\mathrm{Imp})} & = & k_1 F_{\mu\nu}(L^{(1,1)}) + k_2 F_{\mu\nu}(L^{(2,2)}) \nonumber \\
               &   & + k_3 F_{\mu\nu}(L^{(1,2)}) + k_4 F_{\mu\nu}(L^{(1,3)}) \nonumber \\
               &   & + k_5 F_{\mu\nu}(L^{(3,3)})
\end{eqnarray}
where the $k_1,...,k_5$ are improvement factors which take the values 
\begin{eqnarray}
k_1 & = & 19/9 - 55 k_5 \nonumber \\
k_2 & = & 1/36 - 16 k_5 \nonumber \\
k_3 & = & 64 k_5 - 32/45 \nonumber \\
k_4 & = & 1/15 - 6 k_5 \nonumber
\end{eqnarray}
and $k_5$ is a free parameter, which enables us to create 3-loop, 4-loop, 
and 5-loop improved definitions of the field-strength, all of which are ${\cO}(a^4)$-improved.\\
In addition to constructing $Q$, the field strength may be used in (\ref{eq:WilsonAction})
to create a ``reconstructed action''. Since this action is improved in a different manner 
to the cooling action, comparison of both actions may serve as a mechanism 
with which to test the improvement procedures.\\
In this brief report we will simply ``test the water'' with the values \mbox{$c_5=k_5=1/20$} 
(5-loop), $c_5=k_5=0$ (4-loop), and $c_5=1/10$, $k_5=1/90$ (3-loop), and show that we obtain 
excellent results even though many other 5-loop operators, corresponding to other choices of 
$c_5$ and $k_5$, are possible.

\section{RESULTS AND COMPARISON}

\subsection{Constructing Configurations} 
The configurations used in our investigations are generated using the Cabibbo-Marinari 
pseudo-heatbath algorithm \cite{CabMar}. We generate $SU(3)$ configurations by looping twice
over three diagonal $SU(2)$ subgroups. From a cold start (all link values set to zero) we 
thermalize for 5000 sweeps using an ${\cal O}(a^2)$-improved action and a fixed mean-link value, 
and then select configurations every 500 sweeps to create an ensemble of 
configurations with non-trivial topology,  
which are statistically distinct from each other.\\
To reveal the semi-classical structure of the configurations we use a cooling 
algorithm which likewise uses diagonally embedded $SU(2)$ subgroups, with appropriate link 
partitioning \cite{Adel1}.\\
The results we present are produced from configurations on a $12^3\times~24$ lattice with 
$\beta=4.60$, providing a lattice spacing of $a=0.125$ fm.\\

\begin{figure}
\includegraphics[angle=90,width=15pc]{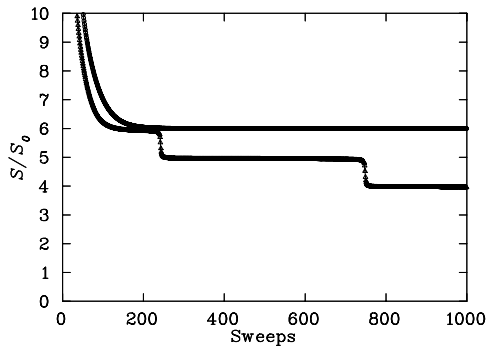}
\caption{Action obtained with 1-loop ($\vartriangle$) and 2-loop ($\circ$) cooling.}
\label{fig:1Land2L}
\end{figure}

\subsection{Comparison of Cooling Schemes}
In practice, because of the large discretization errors present, a cooling scheme based upon 
the plaquette action will destroy topological structure over the whole lattice if it proceeds 
for long enough. However, a cooling scheme using an improved action will stabilise at a 
state with non-trivial structure. We would anticipate cooling schemes with small 
discretization errors will reproduce expected continuum behaviour, especially integer values of
$Q$ and $S/S_0$, and remain stable for hundreds of cooling sweeps. When comparing several 
cooling schemes, the one which fulfills these criteria best will be the preferred scheme for 
studying gauge field dynamics.\\
\begin{figure}
\includegraphics[angle=90,width=15pc]{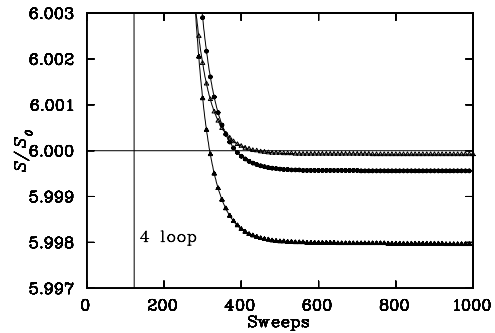}
\caption{Action obtained with 2-loop ($\blacktriangle$), 3LIS ($\vartriangle$), 4LIS (marked), 
and 5LIS ($\bullet$) cooling.}
\label{fig:manyloopaction}
\end{figure}
In Figure~\ref{fig:1Land2L} we can see how the use of a 2-loop, ${\cO}(a^2)$-improved action 
in the cooling algorithm stabilises a configuration at a state with $S/S_0 \approx 6$, 
indicating the presence of six instantons and/or anti-instantons, while plaquette-based cooling
gradually destroys the structure of the configuration.\\ 
In Figure~\ref{fig:manyloopaction} we compare the values of the action obtained with 2-loop, 3LIS,
4LIS, and 5LIS-based cooling on the same configuration used in Figure~\ref{fig:1Land2L}. We can 
see that the 4LIS curve drops well below the desired integer value of six, indicating the 
presence of large negative discretization errors. This comparison and similar work with other 
configurations have lead us to conclude that 3LIS-based cooling gives the most integer-like 
results in general, and furthermore it is faster than 4LIS or 5LIS-based cooling, since it 
requires fewer computational steps.

\subsection{Comparison of Topological Charge Improvement}

\begin{figure}
\includegraphics[angle=90,width=15pc]{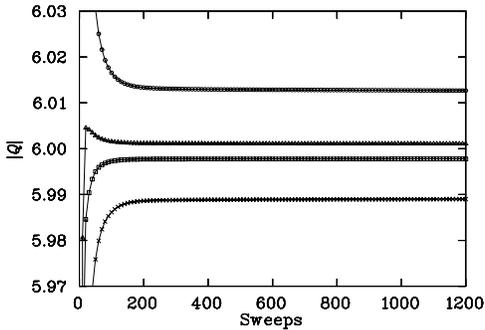}
\caption{$|Q|$ of 2L ($\times$), 3L ($\vartriangle$), 4L ($\square$), and 5L ($\circ$), 
with 3LIS cooling}
\label{fig:manyloopsTQ}
\end{figure}

In Figure~\ref{fig:manyloopsTQ} we compare the various improved $Q$ values obtained as we cool a 
configuration. The 3-loop improved operator gives the most continuum-like results, coming 
to within around 1 part in 6000 of an integer, and has the smallest computational cost of 
all the ${\cO}(a^4)$-improved operators. Similar comparisons with other configurations have 
lead us to select the 3-loop operator as our preferred choice for studying the topological 
charge of cooled configurations. 
 
\subsection{Investigation of the Nahm Transform}
\begin{figure}
\includegraphics[angle=90,width=15pc]{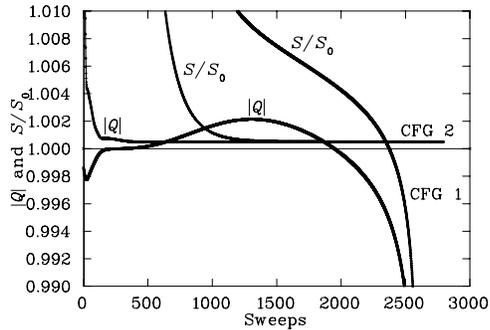}
\caption{3-loop improved $|Q|$ and reconstructed $S/S_0$ of CFG~1 and CFG~2}
\label{fig:CompareQeq1and2}
\end{figure}

Having established our preferred operators for investigating instanton dynamics, we are now in a 
position to compare the stability of $|Q|=1$ and $|Q| \neq 1$ configurations, to determine if 
we can find any evidence that self-dual $|Q|=1$ configurations are not permitted on the 4-torus, 
as implied by the Nahm transform.
Figure~\ref{fig:CompareQeq1and2} shows an overlay of the reconstructed action and topological charge 
of CFG~1, a $|Q| \approx 1$ configuration, and the corresponding values for CFG~2, a $|Q| \approx 2$ 
configuration (shifted down by an increment of one unit so that they overlap with those of CFG~1 
on the scale of this diagram). These configurations are separated by 8000 thermalization sweeps. 
We can clearly see that the $|Q| \approx 2$ configuration stabilises quickly at a self-dual 
near-integer value, while the other configuration does not achieve self-duality and eventually 
destabilises. Investigations of other configurations have shown equivalent behaviour. 

\section{CONCLUSIONS}
Highly improved operators enable us to cool configurations to stable non-trivial self-duality. 
Counting the number of instantons and anti-instantons present by evaluating $S/S_0$ and $Q$ 
allows us to detect evidence of continuum-like behaviour. The long-term stability of 
$|Q| \approx 2$ configurations indicates that our cooling algorithms are highly reliable, 
and hence the destabilization of $|Q| \approx 1$ configurations is due to the consequences 
of the Nahm transform, indicating that on the lattice, just as in the continuum, a configuration 
may not {\em simultaneously} fulfill the two criteria $S/S_0=1$ and $|Q|=1$ on the 4-torus.

\end{document}